\documentclass[aps,prd,preprintnumbers,superscriptaddress,nofootinbib,notitlepage]{revtex4-2}
\usepackage[pdftex]{graphicx}
\usepackage{bm,latexsym,amsmath,amssymb,amsfonts,mathrsfs}
\usepackage{mathtools}
\usepackage{color}
\allowdisplaybreaks[1]
\usepackage[pdftex,colorlinks=true,linkcolor=blue,citecolor=cyan]{hyperref}
\newcommand*{\D}{{\rm d}}

\begin{document}
%-----------------------------------------------------------------%
\title{Black hole perturbations in spatially covariant gravity with just two tensorial degrees of freedom}
%-----------------------------------------------------------------%
\author{Jin~Saito}
\email[Email: ]{j\_saito@rikkyo.ac.jp}
\affiliation{Department of Physics, Rikkyo University, Toshima, Tokyo 171-8501, Japan
}
\author{Tsutomu~Kobayashi}
\email[Email: ]{tsutomu@rikkyo.ac.jp}
\affiliation{Department of Physics, Rikkyo University, Toshima, Tokyo 171-8501, Japan
}
%-----------------------------------------------------------------%
\begin{abstract}
We study linear perturbations around a static and spherically symmetric black hole solution in spatially covariant gravity with just two tensorial degrees of freedom.
In this theory, gravity modification is characterized by a single time-dependent function that appears in the coefficient of $K^2$ in the action, where $K$ is the trace of the extrinsic curvature.
The background black hole solution is given by the Schwarzschild solution foliated by the maximal slices and has a universal horizon at which the lapse function vanishes.
We show that the quadratic action for the odd-parity perturbations is identical to that in general relativity upon performing an appropriate coordinate transformation.
This in particular implies that the odd-parity perturbations propagate at the speed of light, with the inner boundary being the usual event horizon.
We also derive the quadratic action for even-parity perturbations.
In the even-parity sector, one of the two tensorial degrees of freedom is mixed with an instantaneous scalar mode, rendering the system distinct from that in general relativity.
We find that monopole and dipole perturbations, which are composed solely of the instantaneous scalar mode, have no solutions regular both at the universal horizon and infinity (except for the trivial one corresponding to the constant shift of the mass parameter).
We also consider stationary perturbations with higher multipoles.
By carefully treating the locations of the inner boundary, we show that also in this case there are no solutions regular both at the inner boundary and infinity.
Thus, the black hole solution we consider is shown to be perturbatively unique.
\end{abstract}
%-----------------------------------------------------------------%
\preprint{RUP-23-14}
\maketitle
%-----------------------------------------------------------------%
\section{Introduction}

Lovelock's theorem~\cite{Lovelock:1971yv,Lovelock:1972vz} states that a diffeomorphism-invariant theory constructed only from the metric tensor in four dimensions leads uniquely to the Einstein equations in general relativity (GR).
To modify gravity, one must therefore break at least one of the postulates of the theorem,
A typical way of modifying gravity is to add a new dynamical degree of freedom (DOF).
The simplest example is scalar-tensor gravity having one scalar and two tensorial DOFs,
which has been studied extensively, with a particular focus over the past decade on the Horndeski theory~\cite{Horndeski:1974wa,Deffayet:2011gz,Kobayashi:2011nu} and its extensions called degenerate higher-order scalar-tensor (DHOST) theories~\cite{Zumalacarregui:2013pma,Gleyzes:2014dya,Langlois:2015cwa,Crisostomi:2016czh,BenAchour:2016fzp,BenAchour:2016cay} (see Ref.~\cite{Kobayashi:2019hrl} for a review).
An apparently different approach is abandoning full diffeomorphism invariance.
For example, one can consider spatially covariant theories respecting only spatial diffeomorphism invariance under the transformations of spatial coordinates, $x^i\to \tilde x^i(t,x^j)$~\cite{Gao:2014soa,Gao:2014fra}.
However, this approach is basically equivalent to adding new dynamical degrees of freedom and spatially covariant gravity is regarded as a gauge-fixed version of a fully covariant scalar-tensor theory.
; Indeed, starting from a spatially covariant theory, one can introduce a St\"{u}ckelberg scalar field to restore full diffeomorphism invariance and write an action for a corresponding fully covariant scalar-tensor theory~\cite{Gao:2020yzr}.

%~\cite{Gao:2014fra,Fujita:2015ymn,Gao:2018znj,Gao:2019lpz}.

In scalar-tensor theories, the dynamical scalar DOF (say $\phi$) obeys a wave-like equation with some propagation speed $c_s$, the concrete form of which depends on the Lagrangian.
An interesting twist is a case where $\phi$ is an instantaneous mode, $c_s=\infty$, which occurs when the form of the Lagrangian is chosen appropriately.
In such theories, only tensorial DOFs propagate, while the configuration of the instantaneous scalar DOF is determined completely from boundary conditions.
Within the so-called $P(\phi,X)$ theory (where $X=-(\partial\phi)^2/2$), the cuscuton theory, $P=\mu(\phi)\sqrt{X}-V(\phi)$, gives rise to such a nonpropagating scalar field~\cite{Afshordi:2006ad}, leading to interesting cosmology~\cite{Afshordi:2007yx}.
The cuscuton theory was extended in Ref.~\cite{Iyonaga:2018vnu} by demanding that $c_s=\infty$ in the Horndeski and slightly more general theories.
Performing a more rigorous Hamiltonian analysis, one can determine the subset of spatially covariant theories of gravity having two tensorial DOFs only.
This was done in the case where the action is linear in the lapse function in Ref.~\cite{Lin:2017oow} and then the general conditions to render the scalar mode nondynamical were derived in Ref.~\cite{Gao:2019twq}.
Spatially covariant gravity satisfying these conditions
is dubbed as a TTDOF theory.
Since the dynamical DOFs in TTDOF theories are the same as those in GR,
TTDOF gravity may be thought of as minimal modification of GR.
Determining the most general form of the action for TTDOF theories has turned out to be very difficult, and the authors of Ref.~\cite{Gao:2019twq} managed to obtain a particular example of a family of TTDOF theories assuming that the action is quadratic in the extrinsic curvature and linear in the intrinsic curvature as in the Arnowitt-Deser-Misner (ADM) expression for the Einstein-Hilbert term.
The resultant action contains that for the extended cuscuton theory as a special case.

A family of TTDOF theories obtained in Ref.~\cite{Gao:2019twq} has several time-dependent parameters, which can in principle be freely chosen from a purely theoretical point of view.
Some of them can however be fixed by requiring that the speed of gravitational waves is equal to that of light and
the usual GR behavior is restored in the weak-gravity regime in the Solar System~\cite{Iyonaga:2021yfv}.
The TTDOF theory with the thus reduced parameter space
can still mimic the background evolution of the standard $\Lambda$CDM model~\cite{Hiramatsu:2022ahs}.
Interestingly, Ref.~\cite{Hiramatsu:2022ahs} studied the CMB constraints on the TTDOF cosmological model and reported a $\sim$ 4$\sigma$ deviation from the $\Lambda$CDM model based on GR.

Ignoring the terms that are relevant only on cosmological scales, we have a TTDOF theory characterized by a single time-dependent parameter, the impacts of which are not seen in the propagation of gravitational waves and in the weak-gravity regime in the Solar System.
The theory admits the Schwarzschild solution foliated by the maximal slices~\cite{Iyonaga:2021yfv}.
This can easily be seen by noting that the modified gravity parameter enters only in the coefficient of the $K^2$ term in the TTDOF action we are considering, where $K$ is the trace of the extrinsic curvature.
The situation here is quite similar to that of Einstein-Aether theory and the low-energy limit of Horava gravity~\cite{Barausse:2011pu,Blas:2011ni}.
Such a black hole solution in modified gravity is sometimes called stealth, as the geometry is described by the Schwarzschild solution in GR even though the St\"{u}ckelberg scalar exhibits a nontrivial configuration.
In this paper, we study linear perturbations around this stealth black hole solution in the TTDOF theory, paying particular attention to the behavior of the instantaneous scalar mode whose inner boundary conditions are imposed at the universal horizon (the location at which the lapse function vanishes) rather than the usual event horizon.
One of our goals is to see whether the stealth Schwarzschild solution is unique or not.
The question was partially addressed in Ref.~\cite{Iyonaga:2021yfv},
but there only static monopole perturbations were considered.
In the present paper, we start with deriving quadratic actions for all (time-dependent) perturbations with higher multipoles, following closely the formulation of black hole perturbation theory in Horndeski~\cite{Kobayashi:2012kh,Kobayashi:2014wsa} and DHOST gravity~\cite{Takahashi:2019oxz,Tomikawa:2021pca,Takahashi:2021bml}
(see also Refs.~\cite{deRham:2019gha,Khoury:2020aya,Langlois:2021aji,Langlois:2022ulw,Nakashi:2022wdg}).
This result itself has wider applications than merely investigating perturbative uniqueness of the background black hole solution.
See also Ref.~\cite{Lin:2017jvc} for a related analysis of the behavior of the instantaneous scalar field around a black hole with a universal horizon.

This paper is organized as follows. In the next section, we briefly review the TTDOF theory and its static and spherically symmetric black hole solution.
In Sec.~\ref{odd-parity}, we study the odd-parity perturbations.
Then, we consider the even-parity perturbations, presenting the main results separately for monopole, dipole, and higher-multipole modes in Sec.~\ref{even-parity}.
Finally, we draw our conclusions in Sec.~\ref{conclusion}.

\section{A black hole solution in TTDOF theory}

In this section, we introduce the TTDOF theory~\cite{Gao:2019twq} and its static and spherically symmetric black hole solution~\cite{Iyonaga:2021yfv}.

%at rest with respect to the preferred frame.

\subsection{TTDOF theory}

We consider modified gravity with just two tensorial degrees of freedom (TTDOF) developed in Ref.~\cite{Gao:2019twq}.
The symmetry of the theory is the spatial diffeomorphism invariance under
$x^i\to \tilde x^i(t,x^j)$, and therefore we use the ADM variables
to write the action. %on constant time hypersurface.
Specifically, the action that we consider in this paper is given by
\begin{align}
    S=\frac{1}{2}\int \D t\D^3 x \sqrt{\gamma}N\left[K_{ij}K^{ij}-\frac{1}{3}\left(\frac{2N}{\beta+N}+1\right)K^2+R\right], \label{eq:Lag_full}
\end{align}
where $N$ is the lapse function,
$R$ is the three-dimensional Ricci scalar calculated from the spatial metric $\gamma_{ij}$, and
$K_{ij}$ is the extrinsic curvature of constant time hypersurfaces,
\begin{align}
    K_{ij}:=\frac{1}{2N}\left(\partial_t \gamma_{ij}-D_i N_j-D_j N_i\right),
\end{align}
with $N_i$ being the shift vector and $D_i$ the covariant derivative operator associated with $\gamma_{ij}$.
Here, $\beta=\beta(t)$ is an arbitrary function of time characterizing a modification of GR,
and by setting $\beta=0$ the action~\eqref{eq:Lag_full}
reduces to the ADM expression of the Einstein-Hilbert action.
Among a large family of TTDOF theories introduced in Ref.~\cite{Gao:2019twq},
we focus on its particular subset described by the action~\eqref{eq:Lag_full},
because standard gravity is reproduced in the solar system
(in the sense that the parameterized post-Newtonian parameter $\gamma=1$)
and gravitational waves propagate at the speed of light~\cite{Iyonaga:2021yfv},
while rendering the cosmology nontrivial in this theory~\cite{Hiramatsu:2022ahs}.

One can recover the full four-dimensional diffeomorphism invariance
by introducing the St\"{u}ckelberg scalar field.
The resultant covariant expression of the action~\eqref{eq:Lag_full}
belongs to the so-called U-degenerate theory~\cite{Iyonaga:2021yfv},
which is a higher-order scalar-tensor theory satisfying
the degeneracy conditions only when one takes the unitary gauge~\cite{DeFelice:2018ewo}.
However, the apparent scalar degree of freedom is in fact an instantaneous mode having infinite propagation speed and obeying an elliptic equation.
Therefore, the scalar field does not propagate and its behavior is determined completely by boundary conditions.

\subsection{A black hole solution}

The ADM variables for a static and spherically symmetric solution are of the form
\begin{align}
    N=N(r),\quad N_i\D x^i=B(r)F(r)\D r,\quad \gamma_{ij}\D x^i\D x^j =
    F^2\D r^2+r^2\sigma_{ab}\D x^a\D x^b\label{metric:background},
\end{align}
with $\sigma_{ab}\D x^a\D x^b=\D\theta^2+\sin^2\theta\D\varphi^2$ ($a,b=\theta,\varphi$).
Note at this point that we cannot set $B=0$ in general when working in the ADM action (i.e., in the unitary gauge),
because we no longer have the freedom to perform a coordinate transformation $t\to \tilde t(t,r)$.

Substituting Eq.~\eqref{metric:background} to the action~\eqref{eq:Lag_full} and
varying it with respect to $N$, $B$, and $F$, we obtain a set of the field equations
for these variables.
A general solution has not been derived, but one can at least find
the following particular solution:
\begin{align}
    N=N_0\sqrt{f(r)},
    \quad 
    F(r)=\frac{1}{\sqrt{f(r)}},
    \quad 
    B(r)=\frac{N_0 b_0}{r^2},
\end{align}
with 
\begin{align}
    f(r):=1-\frac{\mu_0}{r}+\frac{b_0^2}{r^4},
\end{align}
where $N_0$, $\mu_0$, and $b_0$ are integration constants. 
The solution represents a foliation of the Schwarzschild geometry by maximal slices ($K=0$).
Indeed, by performing the coordinate transformation
\begin{align}
    \tau =N_0 t - \int^r\frac{b_0/r^2}{\sqrt{f}(1-\mu_0/r)}\D r,
    \label{Sch-coord}
\end{align}
and introducing the St\"{u}ckelberg field $\phi=t(\tau, r)$ accordingly,
one obtains $\D s^2=-(1-\mu_0/r)\D \tau^2+(1-\mu_0/r)^{-1}\D r^2+r^2\sigma_{ab}\D x^a\D x^b$,
which is nothing but the standard Schwarzschild metric.
From this observation, we see that $\mu_0$ is the mass parameter, $b_0$ characterizes the foliation, and $N_0$ just corresponds to the rescaling of the time coordinate.
It should be emphasized that the TTDOF theory with the action~\eqref{eq:Lag_full}
admits the above solution even if $\beta$ is an arbitrary time-dependent function.

\begin{figure}[tb]
                \begin{center}
                        \includegraphics[keepaspectratio=true,height=50mm]{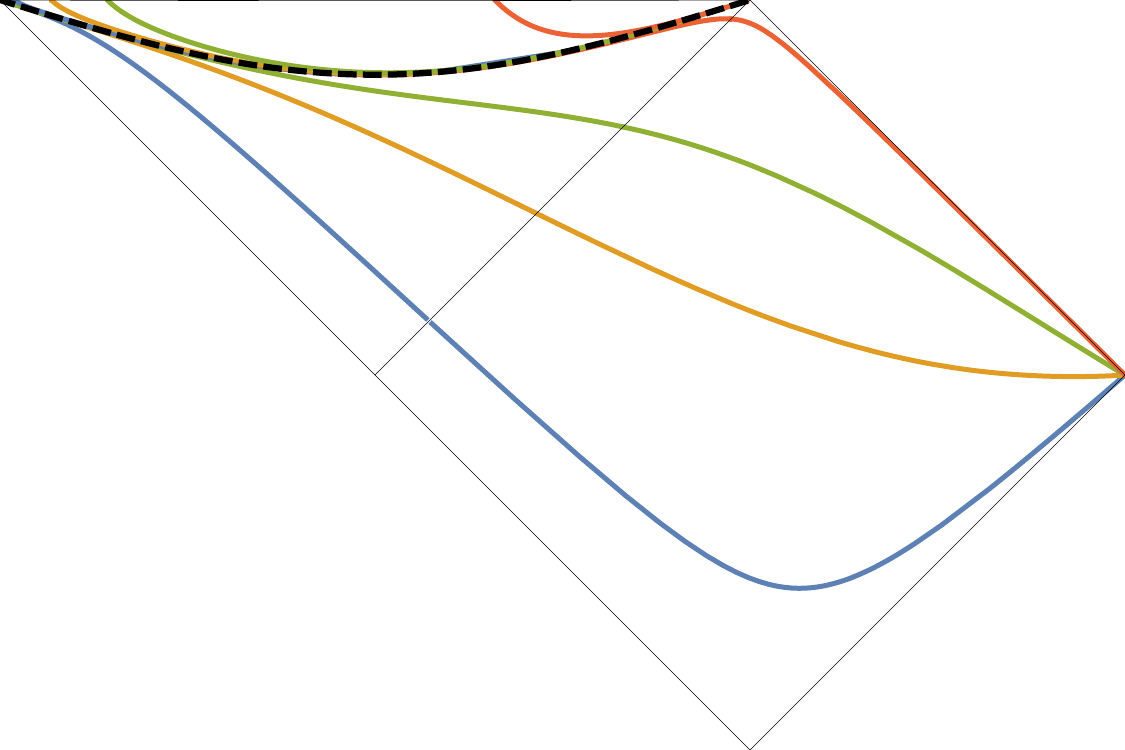}
                \end{center}
                        \caption{Colored lines represent $t=\,$const hypersurfaces,
                        and the black dashed line shows the location of the universal horizon,
                        $r=3\mu_0/4$.}
                \label{fig:conf-diag.pdf}
        \end{figure}

For $b_0\le b_{0,c}:=3\sqrt{3}\mu_0^2/16$, $N(r)$ vanishes at some $r\,(>0)$.
The location at which $N(r)=0$ is called the universal horizon,
which is the causal boundary for the scalar mode with infinite propagation speed.
Only for $b_0=b_{0,c}$ the universal horizon is regular,
while $f(r)$ changes its sign at the universal horizon for $b_0<b_{0,c}$.
In the rest of the paper, we basically consider the black hole solution with
a regular universal horizon (though the perturbation equations can be derived
without regard to the value of $b_0$).
For $b_0=b_{0,c}$, one has
\begin{align}
    f(r)=\left(1-\frac{3\mu_0}{4r}\right)^2\left(1+\frac{\mu_0}{2r}+\frac{3\mu_0^2}{16r^2}\right)\ge 0,
\end{align}
and the universal horizon is located at $r=3\mu_0/4$.
Constant $t$ hypersurfaces and the regular universal horizon are depicted in the Penrose diagram of the Schwarzschild geometry in Fig.~\ref{fig:conf-diag.pdf}.
For $b_0>b_{0,c}$, $f(r)$ is positive everywhere, and therefore the instantaneous mode
is accessible to the singularity at $r=0$. We do not consider this case in this paper.

%If $b_0<b_{0,\mathrm{c}}$, there are regions where $f(r)<0$. In these regions, $N(r)$ becomes a complex number so this case is nonphysical.

%----------------------------------------------%
%-----------------------------------------------%

Here it should be noted again that gravitational waves propagate at the speed of light in our TTDOF theory.
While the causal boundary for the instantaneous scalar mode is given by the universal horizon at $r=3\mu_0/4$, that for gravitational waves is still the usual event horizon at $r=\mu_0$.

In the rest of the paper, we set $\mu_0=1$ to simplify the expressions.

\subsection{A test scalar field on the black hole background}

We are going to discuss the metric perturbations on the black hole background introduced above,
which are composed of the usual tensorial degrees of freedom and the instantaneous scalar mode.
Since the system of the perturbation equations for the metric is highly involved,
it is instructive to consider here a test scalar field $\pi(t,x^i)$ obeying an elliptic equation on
constant $t$ hypersurfaces as the simplest model of an instantaneous mode:
\begin{align}
    S_\pi:=-\frac{1}{2}\int\D t\D^3x \sqrt{\gamma}N\nabla_i\pi\nabla^i\pi. 
\end{align}
This example helps us to see what will be done in the next two sections.

For the above black hole background, the equation of motion for $\pi$ reads
\begin{align}
    f\pi''+f'\pi'-\left[\frac{\ell(\ell+1)}{r^2}+\frac{f'}{r}\right]\pi=0,
\end{align}
where a prime stands for differentiation with respect to $r$ and
the angular part of the Laplacian was replaced by the eigenvalue $-\ell(\ell+1)/r^2$.
The natural inner boundary is seen to be the universal horizon at which $f=0$.
In the case of $\ell=0$, it is easy to find the analytic solution
\begin{align}
    \pi(t,r) = r\left[c_1(t)+c_2(t)\int_r^\infty\frac{\D r}{r^2f}\right],
\end{align}
where $c_1$ and $c_2$ are integration functions.
For large $r$ we have
\begin{align}
    \pi\simeq c_1 r+c_2\left(1+\frac{1}{2r}\right),
\end{align}
while in the vicinity of the universal horizon, we have 
\begin{align}
    \pi\simeq \frac{3c_2}{8(r-3/4)}-\frac{c_2}{3}\ln(r-3/4).
\end{align}
In order for $\pi$ not to diverge at infinity, we require that $c_1=0$.
However, if $c_2\neq 0$ then $\pi$ diverges at the universal horizon.
Therefore, the only possible regular solution is given by $\pi=0$.

We will perform a similar analysis for metric perturbations.
In the case of metric perturbations, multiple components of the metric are coupled,
leading to a more complicated system.
Moreover, the system contains higher spatial derivatives
(for even-parity perturbations with $\ell\ge 1$), allowing for more linearly independent solutions.

\section{Odd-parity perturbations}\label{odd-parity}

In this section and the next, we study linear perturbations around the black hole solution introduced in the previous section.
We first consider odd-parity perturbations in this section.
Only the gravitational-wave degrees of freedom participate in the odd-parity sector.
Noting that the TTDOF theory~\eqref{eq:Lag_full} is built so that gravitational waves propagate in the same way as in GR, we expect that the odd-parity sector is identical to that in GR,
except that the background metric is written in the nonstandard coordinate system.
Indeed, a modification from GR appears only in the $K^2$ term in action,
and it is easy to see that
this term does not include any contribution from odd-parity perturbations.
Below we will see this more explicitly.

Using the spherical harmonics $Y_{\ell m}(\theta,\varphi)$,
we expand the odd-parity perturbations as
\begin{align}
    \delta N_a &= \sum_{\ell,m}h_0^{(\ell m)}(t,r)
    \epsilon^b_{~a}\partial_bY_{\ell m},
    \\
    \delta \gamma_{ra} &= \sum_{\ell,m}h_1^{(\ell m)}(t,r)
    \epsilon^b_{~a}\partial_bY_{\ell m},
    \\
    \delta \gamma_{ab}&=\frac{1}{2}
    \sum_{\ell,m}h_2^{(\ell m)}(t,r)
    \left[\epsilon_a^{~c}\nabla_c\nabla_b+\epsilon_b^{~c}\nabla_c\nabla_a\right]Y_{\ell m},
\end{align}
where $\nabla_a$ is the covariant derivative defined with respect to $\sigma_{ab}$
and $\epsilon_{ab}$ is the Levi-Civita tensor with $\epsilon_{\theta\varphi}=\sin\theta$
and $\epsilon_{\theta}^{~\varphi}=\epsilon_{\theta a}\sigma^{a\varphi}=1/\sin\theta$.
Note that $h_2^{(\ell m)}=0$ for $\ell=1$.
Under a gauge transformation
\begin{align}
x^a\to x^a+\xi^a,\quad \xi_a:=\sum_{\ell,m}\xi^{(\ell m)}(t,r)
\epsilon^b_{~a}\partial_bY_{\ell m},
\end{align}
$h_2^{(\ell m)}$ transforms as
\begin{align}
h_2^{(\ell m)}\to h_2^{(\ell m)}+2r^2\xi^{(\ell m)}.
\end{align}
We thus choose to remove $h_2^{(\ell m)}$ for $\ell\ge 2$.

The quadratic action for the odd-parity perturbations can be written in the form
\begin{align}
    S^{\textrm{(odd)}}=\sum_{\ell,m}\int\D t\D r\mathcal{L}_{\ell m}^{\textrm{(odd)}}.
\end{align}
Let us consider the quadratic Lagrangian with $\ell\ge 2$
(the $\ell=1$ sector must be treated separately).
Thanks to the symmetry, it is sufficient to study the Lagrangian with $m=0$:
\begin{align}
    \mathcal{L}_{\ell\ge2,m=0}^{\textrm{(odd)}}&=
    %\frac{1}{4}\left[\frac{2j^2}{r^2}\left(\frac{r}{FN}\right)'+\frac{c_\ell F}{r^2N}\right]h_0^2
    %-\frac{c_\ell}{4}\left(\frac{N^2-B^2}{r^2FN}\right)h_1^2
    %+\frac{j^2}{4FN}\left[
    %\left(\dot h_1-h_0'\right)^2+\frac{4}{r}h_0\dot h_1
    %\right]-\frac{c_\ell}{2}\frac{B}{r^2N}h_0h_1,
    \frac{1}{4N_0r^2}\left[2j^2+\frac{c_\ell}{f}\right]h_0^2
    -\frac{c_\ell N_0}{4r^2}\left(1-\frac{1}{r}\right)h_1^2
    +\frac{j^2}{4N_0}\left[
    \left(\dot h_1-h_0'\right)^2+\frac{4}{r}h_0\dot h_1
    \right]-\frac{c_\ell}{2}\frac{b_0}{r^4\sqrt{f}}h_0h_1,\label{eq:Lag-odd-1}
\end{align}
where $j^2:=\ell(\ell+1)$ and $c_{\ell}:=(\ell-1)\ell(\ell+1)(\ell+2)$.
Here, a dot and a prime denote derivatives with respect to $t$ and $r$, respectively.
To simplify the expression, we omit the labels $(\ell m)$ from $h_0$ and $h_1$.
We find no terms that depend on $\beta$.
Therefore, the Lagrangian~\eqref{eq:Lag-odd-1} must coincide with that of
the odd-parity perturbations of a Schwarzschild black hole in GR
expressed in the nonstandard coordinate system.
It is obvious that one can arrive at Eq.~\eqref{eq:Lag-odd-1} even if $\beta$ is
a function of time.

To go back to the standard Schwarzschild coordinates,
we use the time coordinate $\tau$ defined in Eq.~\eqref{Sch-coord}
and define
\begin{align}
    \widetilde{h}_0:=\frac{h_0}{N_0},\quad 
    \widetilde{h}_1:=h_1+\frac{b_0/r^2}{\sqrt{f}(1-1/r)}\frac{h_0}{N_0}.
\end{align}
Then, the quadratic action for $\ell\ge 2$ and $m=0$ can be written as 
\begin{align}
    S^{(\textrm{odd})}_{\ell\ge 2,m=0}
    =\frac{1}{4} \int \D\tau \D r\left\{
    \frac{1}{r^2}\left[2j^2+\frac{c_\ell}{1-1/r}\right]\widetilde{h}_0^2
    -\frac{c_\ell}{r^2}\left(1-\frac{1}{r}\right)\widetilde{h}_1^2
    +j^2\left[
        \left(\partial_\tau\widetilde{h}_1-\widetilde{h}_0'\right)^2
        +\frac{4}{r}\widetilde{h}_0\partial_\tau \widetilde{h}_1
    \right]\right\}.
\end{align}
This coincides with the quadratic action for the odd-parity perturbations
of the Schwarzschild solution in GR under an appropriate identification of variables.
Obviously, the same conclusion should hold for the dipole perturbation,
which corresponds to a slow rotation of the black hole.

A comment is now in order. 
In Ref.~\cite{Zhang:2022fbz}, quasi-normal frequencies of gravitational perturbations of a similar black hole in Einstein-Aether theory were calculated imposing the inner boundary conditions at the universal horizon. We argue, however, that the odd-parity metric perturbations in the present case obey the same equation as in GR, i.e. the Regge-Wheeler equation, and propagate at the speed of light,
which indicates that the inner boundary must be set at the horizon for photons, i.e. the usual event horizon at $r=1$.

\section{Even-parity perturbations}\label{even-parity}

Let us move to discuss the even-parity perturbations.
In usual scalar-tensor theories in which the scalar field is dynamical, both the gravitational-wave and scalar degrees of freedom take part in the odd-parity sector.
Therefore, even-parity perturbations with $\ell\ge 2$ are composed of a mixture of gravitational waves and the scalar, while monopole and dipole perturbations are composed solely of the scalar degree of freedom.
A similar mixing occurs for $\ell\ge 2$ in a tricky way in the present case where the scalar is an instantaneous mode obeying an elliptic equation.
Monopole and dipole perturbations in the present case are nondynamical and their configurations are determined entirely from boundary conditions.
Below we will see these points in more detail.

We write the even-parity perturbations of the ADM variables as
\begin{align}
\delta N&= N(r) \sum_{\ell,m}\mathsf{H}_0^{(\ell m)}(t,r)Y_{\ell m},
\\
\delta N_r&=\sum_{\ell,m}\mathsf{H}_1^{(\ell m)}(t,r)Y_{\ell m},
\\
\delta\gamma_{rr}&=F^2(r)\sum_{\ell,m}\mathsf{H}_2^{(\ell m)}(t,r)Y_{\ell m},
\\
\delta N_a&=\sum_{\ell,m}\mathsf{b}^{(\ell m)}(t,r)\partial_aY_{\ell m},
\\
\delta \gamma_{ra}&=\sum_{\ell,m}\mathsf{a}^{(\ell m)}(t,r)\partial_aY_{\ell m},
\\
\delta \gamma_{ab}&=r^2\sum_{\ell,m}\mathsf{K}^{(\ell m)}(t,r)\sigma_{ab}Y_{\ell m}
+r^2\sum_{\ell,m}\mathsf{G}^{(\ell m)}(t,r)\nabla_a\nabla_bY_{\ell m}.
\end{align}
Note that $\mathsf{a}^{(0 0)}=\mathsf{b}^{(0 0)}=\mathsf{G}^{(0 0)}=0$. 
Under infinitesimal coordinate transformations
\begin{align}
r\to r+\sum_{\ell, m}\xi_r^{(\ell m)}(t,r)Y_{\ell m},
\quad
x^a \to x^a+\sum_{\ell, m}\xi_\Omega^{(\ell m)}(t,r)\nabla^aY_{\ell m},
\end{align}
the perturbation variables transform as
\begin{align}
&\mathsf{H}_0\to\mathsf{H}_0-\frac{N'}{N}\xi_r,
\quad
\mathsf{H}_1\to\mathsf{H}_1-(BF\xi_r)'-F^2\dot\xi_r,
\quad
\mathsf{H}_2\to\mathsf{H}_2-2\xi_r'-\frac{2F'}{F}\xi_r,
  \notag \\
  &
  \mathsf{b}\to\mathsf{b}-r^2\dot\xi_\Omega-BF\xi_r,\quad 
  \mathsf{a}\to \mathsf{a}-r^2\xi_\Omega'- F^2 \xi_r,\quad
\mathsf{K}\to \mathsf{K}-\frac{2}{r}\xi_r,\quad \mathsf{G}\to \mathsf{G}-2\xi_\Omega,
\label{gauge-tr}
\end{align}
where the labels $(\ell m)$ were omitted.
In the following analysis, we will use these gauge degrees of freedom to remove some of the variables.
Note that we are working in the unitary gauge and hence we do not have the freedom to change the time coordinate.

As in the analysis of the odd-parity perturbations, we calculate the quadratic action,
\begin{align}
    S^{\textrm{(even)}}=\sum_{\ell,m}\int\D t\D r\mathcal{L}_{\ell m}^{\textrm{(even)}},
\end{align}
and study the perturbations with $\ell=0$, $\ell=1$, and $\ell\ge 2$ separately.

\subsection{$\ell=0$}\label{even-parity:l=0}

For the monopole perturbations, we are left with $\mathsf{H}_0,\mathsf{H}_1,\mathsf{H}_2$, and $\mathsf{K}$, where we omit the labels $(00)$.
From the transformation rules~\eqref{gauge-tr} it can be seen that
we can impose the gauge condition $\mathsf{K}=0$.
The quadratic Lagrangian for $\ell=0$ is then given by
\begin{align}
{\cal L}_{\ell=0}^{(\textrm{even})}&=\frac{\beta r^2}{12N_0\left(\beta+N_0\sqrt{f}\right)}
\left[\frac{\dot{\mathsf{H}}_2}{\sqrt{f}}-\frac{2N_0b_0}{r^2}
\widetilde{\mathsf{H}}_0'-\frac{2}{r^2}
\left(r^2\widetilde{\mathsf{H}}_1\right)'
\right]^2-\frac{2b_0}{r}\widetilde{\mathsf{H}}_0'\widetilde{\mathsf{H}}_1
-N_0rf \widetilde{\mathsf{H}}_0'\mathsf{H}_2
-\frac{r}{N_0\sqrt{f}}\mathsf{H}_2\dot{\widetilde{\mathsf{H}}}_1,
\end{align}
where we defined the convenient combinations of the variables as
\begin{align}
\widetilde{\mathsf{H}}_0&:=\mathsf{H}_0+\frac{1}{2}\mathsf{H}_2,
\\
\widetilde{\mathsf{H}}_1&:=\sqrt{f}\mathsf{H}_1-\frac{N_0b_0}{r^2}
(\mathsf{H}_0+\mathsf{H}_2).
\end{align}
The Euler-Lagrange equations for $\widetilde{\mathsf{H}}_0, \widetilde{\mathsf{H}}_1$, and $\mathsf{H}_2$ read
\begin{align}
    %\mathcal E_0&:=
    \left(\frac{2b_0}{r}\widetilde{\mathsf{H}}_1+N_0 rf \mathsf{H}_2+\frac{2N_0 b_0}{r^2}\mathsf{A}\right)'&=0,\label{ell0eq1}\\
    %\mathcal E_1&:=
    -\frac{2b_0}{r}\widetilde{\mathsf{H}}_0'+\frac{r}{N_0\sqrt f}\dot{\mathsf{H}}_2
    +2r^2\left(\frac{\mathsf{A}}{r^2}\right)'&=0,\label{ell0eq2}\\
    %\mathcal E_2&:=
    N_0 rf\widetilde{\mathsf{H}}_0'+\frac{r}{N_0\sqrt f}\dot{\widetilde{\mathsf{H}}}_1+\frac{\dot{\mathsf{A}}}{\sqrt f}&=0\label{ell0eq3},
\end{align}
where $\mathsf{A}$ is defined as
\begin{align}
    \mathsf{A}:=\frac{\beta r^2}{6N_0(\beta+N_0 \sqrt f)}\left[\frac{\dot{\mathsf{H}}_2}{\sqrt{f}}-\frac{2N_0b_0}{r^2}
\widetilde{\mathsf{H}}_0'-\frac{2}{r^2}
\left(r^2\widetilde{\mathsf{H}}_1\right)'\right].\label{def:A}
\end{align}

Equations~\eqref{ell0eq1}--\eqref{ell0eq3} can be solved exactly
(for any $\beta=\beta(t)$) as follows.
First, Eq.~\eqref{ell0eq1} can be integrated to give 
\begin{align}
    %\bar{\mathcal E}_0=
    \frac{2b_0}{r}\widetilde{\mathsf{H}}_1+N_0 rf \mathsf{H}_2+\frac{2N_0 b_0}{r^2}\mathsf{A}
    =C_1(t),\label{int-eq1}
\end{align}
where $C_1(t)$ is a time-dependent integration function.
Using Eqs.~\eqref{ell0eq2},~\eqref{ell0eq3}, and~\eqref{int-eq1}, we obtain
\begin{align}
    2N_0^2r^2f^{3/2}\left(\frac{\mathsf{A}}{r^2}\right)'=-\dot{C}_1\label{eq:A},
\end{align}
which can be integrated to give
\begin{align}
    \mathsf{A}=
    \frac{\dot{C}_1 r^2}{2N_0^2}\int^\infty_r \frac{\D r}{r^2f^{3/2}}+C_2 (t)r^2,
\end{align}
where $C_2(t)$ is another integration function.
We then use Eqs.~\eqref{ell0eq2} and~\eqref{def:A} to get 
\begin{align}
    (r^2\widetilde{\mathsf{H}}_1)'=
    -N_0r^3\left(\frac{\mathsf{A}}{r^2}\right)'
    -\frac{3N_0}{\beta}(\beta+N_0\sqrt{f})\mathsf{A}.\label{eq:dH1}
\end{align}
We impose the boundary conditions $\mathsf{H}_0,\mathsf{H}_1,\mathsf{H}_2\to 0$ at infinity.
Then, 
noting that $\mathsf{A}\approx (\dot C_1/2N_0^2)r+C_2r^2$ for large $r$, we see from Eq.~\eqref{eq:dH1} that $\dot C_1=C_2=0$, i.e., $\mathsf{A}=0$,
though $C_1$ can still be a nonvanishing, time-independent constant.
Thus, we obtain
\begin{align}
    \frac{\widetilde{\mathsf H}_1}{\sqrt{4\pi}}=\frac{N_0 C_0(t)}{r^2},
\end{align}
where $C_0(t)$ is an integration function.
(The factor $1/\sqrt{4\pi}$ comes from $Y_{00}$.)
It follows from Eq.~\eqref{int-eq1} that 
\begin{align}
    \frac{\mathsf H_2}{\sqrt{4\pi}}=\frac{1}{f}\left[\frac{\delta\mu}{r}-\frac{2b_0C_0(t)}{r^4}\right],
\end{align}
where now we write $C_1=\sqrt{4\pi}N_0\delta\mu$ with $\delta \mu$ being
a {\em time-independent} constant. Finally, we have
\begin{align}
    \frac{\widetilde{\mathsf{H}}_0}{\sqrt{4\pi}}
    =\frac{\dot{C}_0}{N_0}\int^\infty_r \frac{\D r}{r^2f^{3/2}}.
\end{align}

All the perturbation variables are regular at the usual event horizon, $r=1$.
Notice, however, that $\widetilde{\mathsf{H}}_0$
diverges as $\widetilde{\mathsf{H}}_0\sim (r-3/4)^{-2}$
at the universal horizon.
To avoid this, we are forced to set $\dot C_0=0$.
Now it is easy to see that the above solution for the monopole perturbations
can be reproduced by perturbing the parameters of the background solution as
$\mu_0\to\mu_0+\delta\mu$ and $b_0\to b_0+C_0$, which shows that
$\delta\mu$ simply corresponds to a shift of the mass parameter
and nonvanishing $C_0$ renders the universal horizon singular by enforcing $b_0\neq b_{0,c}$.
Therefore, we conclude that no nontrivial regular solution for the monopole perturbations exists.

Let us look at the above solution from the viewpoint of the St\"{u}ckelberg field.
All the metric perturbations with the coefficients $C_0$ and $\dot C_0$
can be eliminated by performing the coordinate transformation
\begin{align}
    t\rightarrow T=t-\frac{C_0(t)}{N_0}\int^\infty_r \frac{\D r}{r^2f^{3/2}}.
    \label{tTtr}
\end{align}
We thus see that
the geometry remains Schwarzschild (with the mass parameter $1+\delta\mu$),
while one has the fluctuation of the St\"{u}ckelberg field,
$\phi=T+\delta\phi$, where $\delta\phi$ is the minus of the second term in Eq.~\eqref{tTtr}.
However, $\phi$ is singular at $r=3/4$ unless $C_0(t)=0$,
and the same conclusion follows.

Let us give a comment on the difference between the present analysis and that in Ref.~\cite{Iyonaga:2021yfv}.
In Ref.~\cite{Iyonaga:2021yfv}, the authors investigated static monopole deformations of the same black hole solution in the same theory without paying particular attention to the regularity of the inner boundary, while in the present paper we have started from generic time-dependent monopole perturbations and showed explicitly how any time-dependent deformation is prohibited by the boundary condtions.
The result obtained here is not trivial as the fluctuation of the instantaneous scalar could in principle be time-dependent.

\subsection{$\ell=1$}\label{even-parity:l=1}

In the quadratic Lagrangian for the dipole perturbations,
$\mathsf{K}^{(1m)}$ and $\mathsf{G}^{(1m)}$ appears only through the combination
$\mathsf{K}^{(1m)}-\mathsf{G}^{(1m)}$. We may therefore impose the gauge condition
$\mathsf{K}^{(1m)}-\mathsf{G}^{(1m)}=0$ along with $\mathsf{H}_0^{(1m)}=0$
by choosing $\xi_r^{(1m)}$ and $\xi_\Omega^{(1m)}$ appropriately.
Focusing on the $m=0$ part without loss of generality,
the quadratic Lagrangian for $\ell=1$ is found to be
\begin{align}
{\cal L}_{\ell=1,m=0}^{(\textrm{even})}&=\frac{\beta r^2}{12N_0(\beta+N_0\sqrt{f})}
\left[\frac{\dot{\mathsf{H}}_2}{\sqrt{f}}
-\frac{N_0b_0}{r^2}\mathsf{H}_2'-\frac{2}{r^2}\left(r^2\widetilde{\mathsf{H}}_1\right)'
+\frac{4\widetilde{\mathsf{b}}}{r}
+\frac{6(\beta+N_0\sqrt{f})}{\beta r}
\left(\widetilde{\mathsf{H}}_1-\widetilde{\mathsf{b}}\right)
\right]^2
\notag \\
&\quad
+\frac{r^2}{2N_0}\left[
\dot{\widetilde{\mathsf{a}}}-rN_0b_0\left(\frac{\sqrt{f}}{r^3}\widetilde{\mathsf{a}}\right)'
-r\left(\frac{\sqrt{f}}{r}\widetilde{\mathsf{b}}\right)'
-\frac{\widetilde{\mathsf{H}}_1}{r\sqrt{f}}
\right]^2
-\frac{3\sqrt{f}}{\beta}\left(\widetilde{\mathsf{H}}_1-\widetilde{\mathsf{b}}\right)^2
+\frac{6b_0}{r^2}\widetilde{\mathsf{H}}_1\widetilde{\mathsf{a}}
-\frac{3b_0}{r^2}\mathsf{H}_2\widetilde{\mathsf{b}}
\notag \\ &\quad
+N_0f\widetilde{\mathsf{a}}^2-\frac{N_0}{2}\left[f+(rf)'\right]
\widetilde{\mathsf{a}}\mathsf{H}_2+\frac{N_0}{4}(rf)'\mathsf{H}_2^2\label{Lag:even_l=1},
\end{align}
where to simplify the expression we defined
\begin{align}
\widetilde{\mathsf{a}}:=\frac{\mathsf{a}}{r},
\quad
\widetilde{\mathsf{b}}:=\frac{\mathsf{b}}{r\sqrt{f}}-\frac{N_0b_0}{r^3}\mathsf{a},
\quad
\widetilde{\mathsf{H}}_1:=\sqrt{f}\mathsf{H}_1-\frac{N_0b_0}{r^2}\mathsf{H}_2,
\label{def:tilde}
\end{align}
and omitted the labels $(10)$.

We then introduce new variables $\mathsf{P}$ and $\mathsf{Q}$ and define the Lagrangian $\widetilde{\mathcal L}^{(2)}_{\ell=1,m=0}$ as
\begin{align}
    \widetilde{\mathcal L}^{(2)}_{\ell=1,m=0}&=\mathcal L^{(2)}_{\ell=1,m=0}
    \notag \\ & \quad 
    -\frac{\beta r^2}{12N_0(\beta+N_0\sqrt{f})}
\left[\frac{\dot{\mathsf{H}}_2}{\sqrt{f}}
-\frac{N_0b_0}{r^2}\mathsf{H}_2'-\frac{2}{r^2}\left(r^2\widetilde{\mathsf{H}}_1\right)'
+\frac{4\widetilde{\mathsf{b}}}{r}
+\frac{6(\beta+N_0\sqrt{f})}{\beta r}
\left(\widetilde{\mathsf{H}}_1-\widetilde{\mathsf{b}}\right)-\frac{\beta+N_0\sqrt f}{\beta}\mathsf{Q}
\right]^2
\notag\\
&\quad
+\frac{r^2}{2N_0}\left[
\dot{\widetilde{\mathsf{a}}}-rN_0b_0\left(\frac{\sqrt{f}}{r^3}\widetilde{\mathsf{a}}\right)'
-r\left(\frac{\sqrt{f}}{r}\widetilde{\mathsf{b}}\right)'
-\frac{\widetilde{\mathsf{H}}_1}{r\sqrt{f}}-\mathsf{P}
\right]^2.
\end{align}
This Lagrangian is equivalent to the original one $\mathcal{L}^{(2)}_{\ell=1,m=0}$
because the additional parts vanish upon substituting the solutions to the equations of motion for $\mathsf{P}$ and $\mathsf{Q}$. However, the new Lagrangian $\widetilde{\mathcal L}^{(2)}_{\ell=1,m=0}$ is more useful for our anaysis.
Varying $\widetilde{\mathcal L}^{(2)}_{\ell=1,m=0}$ with respect to $\widetilde{\mathsf{H}}_1$, $\mathsf H_2$, $\widetilde{\mathsf a}$, and $\widetilde{\mathsf b}$, we obtain the equations of motion for these variables, which turn out to be the constraint equations.
One can easily solve them to express $\widetilde{\mathsf{H}}_1$, $\mathsf H_2$, $\widetilde{\mathsf a}$, and $\widetilde{\mathsf b}$ in terms of $\mathsf{P}$, $\mathsf{Q}$, and their first derivatives:
$\widetilde{\mathsf H}_1=(\dots)\mathsf P+(\dots)\mathsf Q+(\dots)\dot{\mathsf{P}}+(\dots)\dot{\mathsf Q}+(\dots)\mathsf P'+(\dots)\mathsf Q'$,
$\mathsf{H}_2,\widetilde{\mathsf{a}},\widetilde{\mathsf{b}}=\dots$,
where the explicit expressions are messy.
Substituting these back to $\widetilde{\mathcal L}^{(2)}_{\ell=1,m=0}$,
we can express it in terms of $\mathsf P$, $\mathsf Q$, and their derivatives.
We then introduce the new variable $\chi:=\mathsf Q+3\sqrt{f} \mathsf P$ and replace $\mathsf Q$ in the Lagrangian by $\chi$. By doing so we can remove all the derivatives acting on $\mathsf{P}$.
The equation of motion for $\mathsf P$ can therefore be solved to express it in terms of $\chi$ and its derivatives. Substituting the solution back to the Lagrangian to remove $\mathsf{P}$ and performing integration by parts, we finally arrive at the reduced Lagrangian for a single master variable $\chi$.
Despite lengthy expressions at each intermediate step, the final form of the reduced Lagrangian is rather simple:
\begin{align}
    \widetilde{\mathcal L}^{(2)}_{\ell=1,m=0}=\frac{\beta(t)}{81N_0^2}
    \left[d_1(r)(\chi'')^2+d_2(r)(\chi')^2+d_3(r)\chi^2\right],
\end{align}
where the coefficients $d_1$, $d_2$, and $d_3$ are independent of $\beta(t)$ and are given explicitly by
\begin{align}
    d_1(r)&=\frac{r^8f^{3/2}}{3},\label{dipole-d1}
    \\ 
    d_2(r)&=-2r^6f^{1/2}\left(2-\frac{1}{r}-\frac{5b_0^2}{r^4}\right),
    \label{diole-d2}
    \\ 
    d_3(r)&=\frac{r^4}{f^{1/2}}\left(\frac{2}{r}-\frac{20b_0^2}{r^4}
    +\frac{21b_0^{2}}{r^5}\right).\label{dipole-d3}
\end{align}
(We are primarily interested in the case where $b_0=b_{0,c}$, but the above expressions are valid for any $b_0$.)
Note that there are no time derivatives of $\chi$ in the Lagrangian, which means that there is no propagating dipole mode.

The equation of motion for $\chi$ is given by the fourth-order differential equation with respect to $r$, 
\begin{align}
    \left(d_1\chi''\right)''-\left(d_2\chi'\right)'+d_3\chi=0,\label{eq:chi}
\end{align}
and the configuration of $\chi$ is determined once one specifies the boundary conditions.
As implied by Eqs.~\eqref{dipole-d1}--\eqref{dipole-d3}, the inner boundary conditions must be imposed at the location at which $f$ vanishes, i.e., the universal horizon.
All the other variables can then be obtained straightforwardly from $\chi$.

We have not found an analytic solution to Eq.~\eqref{eq:chi}.
However, we can obtain solutions valid near the boundaries.
Let us first look for a solution valid for large $r$ in the form
\begin{align}
    \chi=
    \mathcal C_0(t)+\frac{\mathcal C_1(t)}{r}+\frac{\mathcal C_2(t)}{r^2}+\ldots+\log r\left[\mathcal D_0(t)+\frac{\mathcal D_1(t)}{r}+\frac{\mathcal D_2(t)}{r^2}+\dots\right].
    \label{chi-series-inf}
\end{align}
Substituting this into Eq.~\eqref{eq:chi}, one can derive the algebraic relations among the coefficients. One may take $\mathcal C_0$, $\mathcal C_2$, $\mathcal C_3$, and $\mathcal C_5$ as independent coefficients and express all the other coefficients in terms of them.
One can then obtain the metric perturbations
$\mathsf H_1$, $\mathsf H_2$, $\mathsf a$, and $\mathsf b$
in a series expansion similar to Eq.~\eqref{chi-series-inf}.
In order for the metric perturbations to be vanishing at infinity,
it turns out that $\mathcal C_0=\mathcal C_2=\mathcal C_3=0=\ddot{\mathcal{C}}_5$ must be imposed.
Then, $\mathcal{C}_5$ may have the form $\mathcal{C}_5=v_1t+v_0$, where
$v_1$ and $v_0$ are constants, showing that $\chi$ diverges as $t\to\infty$
unless $\dot{\mathcal{C}}_5=v_1=0$.
We thus see that only a \textit{constant} $\mathcal{C}_5$ is allowed to be nonvanishing from the boundary conditions at infinity and $\chi$ is given by
\begin{align}
    \chi=\frac{\mathcal{C}_5}{r^5}\left(1+\frac{3}{4r}+\frac{9}{16r^2}+\frac{7}{16r^3}+\dots\right).\label{bc:chi-inf}
\end{align}
In this way, $\chi$ is uniquely determined up to an overall constant.
The dipole metric perturbations for large $r$ are given accordingly by
\begin{align}
    \widetilde{\mathsf H}_1&=\frac{\mathcal{C}_5}{r^3}\left(\frac{2}{9}+\frac{1}{12}\frac{1}{r}+\frac{1}{24}\frac{1}{r^2}+\dots\right),\\
    \mathsf H_2&=\frac{b_0}{N_0\mu_0}\frac{\mathcal{C}_5}{r^4}\left(\frac{8}{3}+\frac{4}{3}\frac{1}{r}+\frac{5}{6}\frac{1}{r^2}+\dots\right),\\
    \widetilde{\mathsf a}&=\frac{b_0}{N_0}\frac{\mathcal{C}_5}{r^4}\left(\frac{4}{3}+\frac{1}{3}\frac{1}{r}+\dots\right),\\
    \widetilde{\mathsf b}&=-\frac{\mathcal{C}_5}{r^3}\left(\frac{1}{9}+\frac{1}{12}\frac{1}{r}+\frac{1}{16}\frac{1}{r^2}+\dots\right),
\end{align}
where $\beta$ does not appear in the series expansion.

\begin{figure}[tb]            
                    \centering    \includegraphics[keepaspectratio=true,height=50mm]{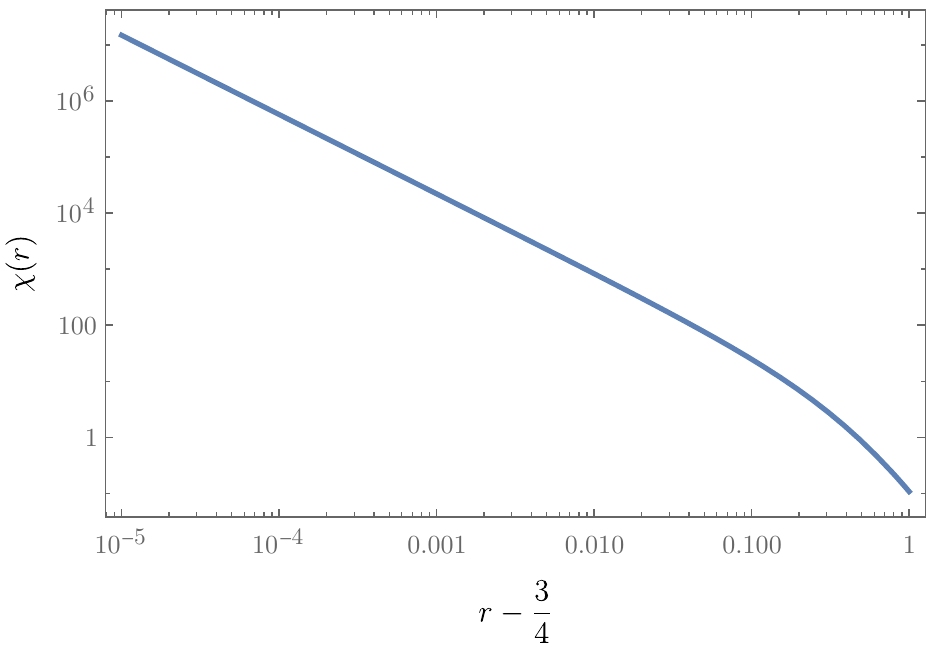}
                 \caption{The behavior of $\chi$ near the universal horizon,
                 obtained by solving Eq.~\eqref{eq:chi} numerically with the boundary condition~\eqref{bc:chi-inf} (with $\mathcal{C}_5=1$).}
                \label{fig:loglog_chi_l_1_nolegends.pdf}
        \end{figure}

Next, let us examine the behavior of the dipole perturbations near the (regular) universal horizon at $r=3/4$, setting now $b_0=b_{0,c}$.
Near the universal horizon, we have
$d_1\simeq (81/256\sqrt{2})\cdot(r-3/4)^3$,
$d_2\simeq (81/256\sqrt{2})\cdot 3(r-3/4)$,
and $d_3\simeq (81/256\sqrt{2})\cdot 4(r-3/4)^{-1}$.
It then follows that, near the universal horizon, $\chi$ is given by a linear combination of the four independent solutions
\begin{align}
    (r-3/4)^{\sqrt{2}},
    \quad (r-3/4)^{\sqrt{2}}\ln (r-3/4),
    \quad (r-3/4)^{-\sqrt{2}},
    \quad (r-3/4)^{-\sqrt{2}}\ln (r-3/4).
\end{align}
By numerically solving Eq.~\eqref{eq:chi} from some large $r$ toward the universal horizon,
we find that the coefficients of the latter two are nonvanishing and $\chi$ diverges,
as can be seen from Fig.~\ref{fig:loglog_chi_l_1_nolegends.pdf}.
This forces the metric perturbations reconstructed from $\chi$ to diverge as
\begin{align}
    (BF)^{-1}\mathsf{H}_1\sim 
    \mathsf{H}_2\sim 
    \mathsf{a}\sim 
    \mathsf{b}\sim (r-3/4)^{-1}\chi.
\end{align}
One can also see that the linear perturbation of the three-dimensional Ricci scalar diverges as $\delta R\sim (r-3/4)^{-1}\chi$.
We, therefore, conclude that no regular dipole perturbations are allowed in the present setup.

%Next, we examine the behavior of the solution near the universal horizon $r=3\mu_0/4$.
%Now, we only consider the case where the regular universal horizon exists: $b_0=b_{0,c}$.
%Solving the equation~\eqref{eq:chi} numerically from a large $r$ towards the universal horizon, the results are shown in Fig.~\ref{fig:num-solution_chi_l=1.pdf}.

% %----------------------------------------------%
% %-----------------------------------------------%

\if0
As seen from Fig.~\ref{fig:loglog_chi_l_1_nolegends.pdf}, we see that the master variable $\chi$ diverges at the universal horizon $r=3\mu_0/4$.
The leading term of the expanding solution of $\chi$ near the universal horizon is given by
\begin{align}
    \chi \approx c(r-3\mu_0/4)^{-\sqrt{2}}
\end{align}
where $c$ is a normalization constant.
Thus the master variable is singular at the universal horizon.
The asymptotic behaviors of the metric perturbations near the universal horizon are also given by
\begin{align}
    \widetilde{\mathsf H}_1&\approx c'(r-3\mu_0/4)^{-1-\sqrt 2},\quad \mathsf{H}_2\approx -\frac{2\sqrt{3}}{N_0}c'(r-3\mu_0/4)^{-1-\sqrt{2}},\\
    \widetilde{\mathsf{a}}&\approx\frac{3\sqrt{3}(\sqrt{2}-1)}{8N_0}c'(r-3\mu_0/4)^{-2-\sqrt{2}},\quad\widetilde{\mathsf{b}}\approx c'(r-3\mu_0/4)^{-1-\sqrt{2}}
\end{align}
where $c'$ is a normalization constant, so the metric perturbations are also singular at the universal horizon.
Therefore there are no dipole perturbations that are regular at both $r\rightarrow \infty$ and the universal horizon $r\rightarrow 3\mu_0/4$.
\fi

\subsection{$\ell\ge 2$}\label{even-parity:l>=2}

Finally, let us consider the even-parity perturbations with $\ell\ge 2$, which contains both gravitational waves and instantaneous mode.
By choosing appropriately $\xi_r^{(\ell m)}$ and $\xi_\Omega^{(\ell m)}$, one can set $\mathsf{K}^{(\ell m)}=\mathsf{G}^{(\ell m)}=0$.
The remaining variables are
$\mathsf{H}_0^{(\ell m)}$, $\mathsf{H}_1^{(\ell m)}$, $\mathsf{H}_2^{(\ell m)}$, $\mathsf{a}^{(\ell m)}$, and $\mathsf{b}^{(\ell m)}$.
Focusing again on the perturbations with $m=0$ and omitting the labels $(\ell m)$, the quadratic Lagrangian is given by
\begin{align}
{\cal L}_{\ell\ge 2,m=0}^{(\textrm{even})}&=\frac{\beta r^2}{12N_0(\beta+N_0\sqrt{f})}
\biggl[\frac{\dot{\mathsf{H}}_2}{\sqrt{f}}
-\frac{N_0b_0}{r^2}\mathsf{H}_2'-\frac{2}{r^2}\left(r^2\widetilde{\mathsf{H}}_1\right)'
+\frac{6(\beta+N_0\sqrt{f})}{\beta r}\widetilde{\mathsf{H}}_1
\notag \\ &\quad
-\frac{6N_0b_0}{\beta r^3}(\beta+N_0\sqrt{f})\mathsf{H}_0
-\frac{j^2(\beta+3N_0\sqrt{f})}{\beta r}\widetilde{\mathsf{b}}
\biggr]^2
%\notag \\&\quad
+\frac{j^2r^2}{4N_0}\left[
\dot{\widetilde{\mathsf{a}}}-rN_0b_0\left(\frac{\sqrt{f}}{r^3}\widetilde{\mathsf{a}}\right)'
-r\left(\frac{\sqrt{f}}{r}\widetilde{\mathsf{b}}\right)'
-\frac{\widetilde{\mathsf{H}}_1}{r\sqrt{f}}
\right]^2
\notag \\ &\quad
-\frac{3\sqrt{f}}{\beta}\left(\widetilde{\mathsf{H}}_1
-\frac{N_0b_0}{r^2}\mathsf{H}_0
-\frac{j^2}{2}
\widetilde{\mathsf{b}}\right)^2
+\frac{3j^2b_0}{r^2}\widetilde{\mathsf{H}}_1\widetilde{\mathsf{a}}
-\frac{3j^2b_0}{2r^2}\mathsf{H}_2\widetilde{\mathsf{b}}
+\frac{j^2}{2}N_0f\widetilde{\mathsf{a}}^2-\frac{j^2N_0}{4}\left[f+(rf)'\right]
\widetilde{\mathsf{a}}\mathsf{H}_2
\notag \\ &\quad
+\frac{N_0}{4}(rf)'\mathsf{H}_2^2
%\notag\\ &\quad
+\frac{(\ell-1)\ell(\ell+1)(\ell+2)}{4N_0}\widetilde{\mathsf{b}}^2+
N_0f\left[
r\mathsf{H}_2'-j^2(r\widetilde{\mathsf{a}})'
\right]\mathsf{H}_0
\notag \\ & \quad
+N_0\left[\frac{j^2}{2}+(rf)'\right]\mathsf{H}_0\mathsf{H}_2
-\frac{j^2}{2}N_0[f+(rf)']\widetilde{\mathsf{a}}\mathsf{H}_0,\label{Lag:ell-2}
\end{align}
where $\widetilde{\mathsf{H}}_1$, $\widetilde{\mathsf{a}}$, and $\widetilde{\mathsf{b}}$ are defined in the same way as in the case of $\ell=1$ (see Eq.~\eqref{def:tilde}) and $j^2:=\ell(\ell+1)$ as before.

Since we only have one of the two tensorial modes as a dynamical degree of freedom in the even-parity sector with $\ell\ge 2$ and there is no dynamical scalar degree of freedom, we expect that by integrating out nondynamical variables in the Lagrangian~\eqref{Lag:ell-2} we would end up with a reduced Lagrangian written in terms of a single master variable.
Unfortunately, however, we have not been able to do so.
Therefore, in what follows we focus on stationary perturbations by dropping all time derivatives and solve directly the equations of motion derived from the Lagrangian~\eqref{Lag:ell-2} without trying to rewrite the system in terms of a single variable.
(In doing so we also assume that $\beta=\,$const.)
This amounts to considering stationary deformations of the black hole while discarding propagating gravitational waves.
Varying the action, one can straightforwardly obtain the equations of motion for $\mathsf{H}_0$, $\widetilde{\mathsf{H}}_1$, $\mathsf{H}_2$, $\widetilde{\mathsf{a}}$, and $\widetilde{\mathsf{b}}$:
\begin{align}
    \frac{\delta S^{\textrm{(even)}}}{\delta \mathsf{H}_0^{(\ell 0)}}=
    \frac{\delta S^{\textrm{(even)}}}{\delta \widetilde{\mathsf{H}}_1^{(\ell 0)}}=
    \frac{\delta S^{\textrm{(even)}}}{\delta \mathsf{H}_2^{(\ell 0)}}=
    \frac{\delta S^{\textrm{(even)}}}{\delta \widetilde{\mathsf{a}}^{(\ell 0)}}=
    \frac{\delta S^{\textrm{(even)}}}{\delta \widetilde{\mathsf{b}}^{(\ell 0)}}=0.
    \label{eq:l=2_even}
\end{align}

First, we determine the behavior of the metric perturbations at large $r$,
assuming the series expansion form
\begin{align}
    \mathsf{H}_0&=r^{-\ell}\left(\mathcal{C}_0^{(0)}+\frac{\mathcal{C}_1^{(0)}}{r}+\frac{\mathcal{C}_2^{(0)}}{r^2}+\dots\right),\quad
    \widetilde{\mathsf{H}}_1=r^{-\ell}\left(\mathcal{C}_0^{(1)}+\frac{\mathcal{C}_1^{(1)}}{r}+\frac{\mathcal{C}_2^{(1)}}{r^2}+\dots\right),\notag \\
    \mathsf{H}_2&=r^{-\ell}\left(\mathcal{C}_0^{(2)}+\frac{\mathcal{C}_1^{(2)}}{r}+\frac{\mathcal{C}_2^{(2)}}{r^2}+\dots\right),\quad 
    \widetilde{\mathsf{a}}=r^{-\ell}\left(\mathcal{C}_0^{(a)}+\frac{\mathcal{C}_1^{(a)}}{r}+\frac{\mathcal{C}_2^{(2)}}{r^2}+\dots\right),\notag \\
    \widetilde{\mathsf{b}}&=r^{-\ell}\left(\mathcal{C}_0^{(b)}+\frac{\mathcal{C}_1^{(b)}}{r}+\frac{\mathcal{C}_2^{(b)}}{r^2}+\dots\right).
\end{align}
Substituting these to the equations of motion~\eqref{eq:l=2_even},
we can derive the algebraic relations among the coefficients $\mathcal{C}_0^{(0)}, \mathcal{C}_0^{(1)}, \dots$.
By inspecting the relations, we find that only two of the coefficients are free and independent, and all the other coefficients are expressed using the two.
Explicitly, we have
\begin{align}
\mathsf{H}_0&=\frac{\mathcal B_0}{r^{\ell+4}}\frac{b_0}{N_0}\left[1+\frac{2\ell^2+5\ell+4}{4(\ell+1)}\frac{1}{r}+\frac{2\ell^3+11\ell^2+21\ell+16}{16(\ell+1)}\frac{1}{r^2}+\dots\right]\notag\\
&\quad-\frac{1}{2(\ell+1)}\frac{\mathcal B_2}{r^{\ell+1}}\left[1+\frac{\ell +2}{2}\frac{1}{r}+\dots\right],\label{sol:H0_l>2}\\
  \widetilde{\mathsf{H}}_1&=\frac{\mathcal B_0}{r^{\ell+2}}\left[1+\frac{\ell(2\ell+1)}{4(\ell+1)}\frac{1}{r}+\frac{\ell(2\ell+1)}{16}\frac{1}{r^2}+\dots\right]\notag\\
  &\quad -\frac{\mathcal B_2}{r^{\ell+3}}N_0 b_0\left[\frac{\ell+2}{4(\ell+1)}+\frac{4\ell^4+35\ell^3+85\ell^2+98\ell+48}{16(\ell+1)(\ell+2)(2\ell+3)}\frac{1}{r}+\dots\right],\label{sol:H1_l>2}\\
  \mathsf H_2&=-\frac{2\mathcal B_0}{r^{\ell+3}}\frac{b_0}{N_0}\left[1+\frac{2\ell^2+5\ell+4}{4(\ell+1)}\frac{1}{r}+\frac{2\ell^3+11\ell^2+21\ell+16}{16(\ell+1)}\frac{1}{r^2}+\dots\right]\notag\\
&\quad +\frac{\mathcal B_2}{r^{\ell+1}}\left[1+\frac{\ell^3+5\ell^2+7\ell+4}{2(\ell+1)(\ell+2)}\frac{1}{r}+\dots\right],\label{sol:H2_l>2}\\
  \widetilde{\mathsf{a}}&=\frac{1}{\ell+1}\frac{\mathcal B_0}{r^{\ell+4}}\frac{b_0}{N_0}\left[1+\frac{2\ell+5}{4}\frac{1}{r}+\frac{2\ell^2+13\ell+22}{16}\frac{1}{r^2}+\dots\right]\notag\\
  &\quad -\frac{\mathcal B_2}{2r^{\ell+1}}\left[\frac{1}{\ell+1}+\frac{\ell+4}{2(\ell+2)}+\dots\right],\label{sol:a_l>2}\\
  \widetilde{\mathsf{b}}&=\frac{1}{\ell+1}\frac{\mathcal B_0}{r^{\ell+2}}\left[1+\frac{2\ell+1}{4}\frac{1}{r}+\frac{(2\ell+1)(\ell+2)}{16}\frac{1}{r^2}+\dots\right]\notag\\
  &\quad -\frac{\mathcal B_2}{r^{\ell+3}}N_0b_0\left[\frac{1}{4(\ell+1)}+\frac{4\ell^2+13\ell+18}{16(\ell+1)(2\ell+3)}\frac{1}{r}+\dots\right],\label{sol:b_l>2}
\end{align}
where we redefined the two independent coefficients and introduced the constants $\mathcal B_0$ and $\mathcal B_2$. The above expressions are valid for any $b_0$.

To see the physical interpretation of the two constants $\mathcal B_0$ and $\mathcal B_2$,
let us perform similar calculations for the Schwarzschild solution,
$\D s^2=-(1-1/r)\D t^2+(1-1/r)^{-1}\D r^2+r^2\sigma_{ab}\D x^a\D x^b$,
in GR.
The quadratic Lagrangian for the even-parity perturbations of the Schwarzschild solution in GR can be reproduced by setting $N_0=1$, $b_0=0$, and $\beta=0$ in the Lagrangian~\eqref{Lag:ell-2} and removing $\mathsf{b}$ with an infinitesimal coordinate transformation of $t$:
\begin{align}
  {\cal L}_{\ell\ge 2,m=0}^{(\textrm{even, GR})}&=r
  \left[\frac{\dot{\mathsf{H}}_2}{\sqrt{f}}
  -\frac{2}{r^2}\left(r^2\widetilde{\mathsf{H}}_1\right)'
  \right]\widetilde{\mathsf{H}}_1
  +\frac{j^2r^2}{4}\left[
  \dot{\widetilde{\mathsf{a}}}
  -\frac{\widetilde{\mathsf{H}}_1}{r\sqrt{f}}
  \right]^2
  +\frac{j^2}{2}f\widetilde{\mathsf{a}}^2-\frac{j^2}{4}\left[f+(rf)'\right]
  \widetilde{\mathsf{a}}\mathsf{H}_2+\frac{1}{4}(rf)'\mathsf{H}_2^2
  \notag\\ &\quad
  +  f\left[
  r\mathsf{H}_2'-j^2(r\widetilde{\mathsf{a}})'
  \right]\mathsf{H}_0
  +\left[\frac{j^2}{2}+(rf)'\right]\mathsf{H}_0\mathsf{H}_2
  -\frac{j^2}{2}[f+(rf)']\widetilde{\mathsf{a}}\mathsf{H}_0,\label{Lag:ell-2gr}
  \end{align}
with $f=1-1/r$.
Similar manipulations as described above lead to
\begin{align}
    \mathsf{H}_0&=-\frac{1}{2(\ell+1)}\frac{\bar{\mathcal{B}}_2}{r^{\ell+1}}\left[1+\frac{\ell+2}{2}\frac{1}{r}+\dots\right],\label{gr-h0}\\
    \widetilde{\mathsf{H}}_1&=0,\\
    \mathsf{H}_2&=\frac{\bar{\mathcal{B}}_2}{r^{\ell+1}}\left[1+\frac{\ell^3+5\ell^2+7\ell+4}{2(\ell+1)(\ell+2)}\frac{1}{r}+\dots\right],\\
    \widetilde{\mathsf{a}}&=-\frac{\bar{\mathcal{B}}_2}{2r^{\ell+1}}\left[\frac{1}{\ell+1}+\frac{\ell+4}{2(\ell+2)}\frac{1}{r}+\dots\right]\label{gr-a},
\end{align}
where it is found that we are allowed to have only one integration constant $\bar{\mathcal{B}}_2$ in the case of GR.
Equations~\eqref{gr-h0}--\eqref{gr-a} agree with Eqs.~\eqref{sol:H0_l>2}--\eqref{sol:b_l>2} with $b_0=0=\mathcal{B}_0$, if one identifies $\mathcal{B}_2$ as $\bar{\mathcal{B}}_2$.
This observation clearly shows that the terms proportional to $\mathcal{B}_2$ come from the tensorial degrees of freedom, while those with $\mathcal{B}_0$ originate from the instantaneous scalar mode.

\begin{figure}[tb]
            \begin{minipage}[b]{0.45\columnwidth}
                %\begin{center}
                    \centering    \includegraphics[keepaspectratio=true,height=50mm]{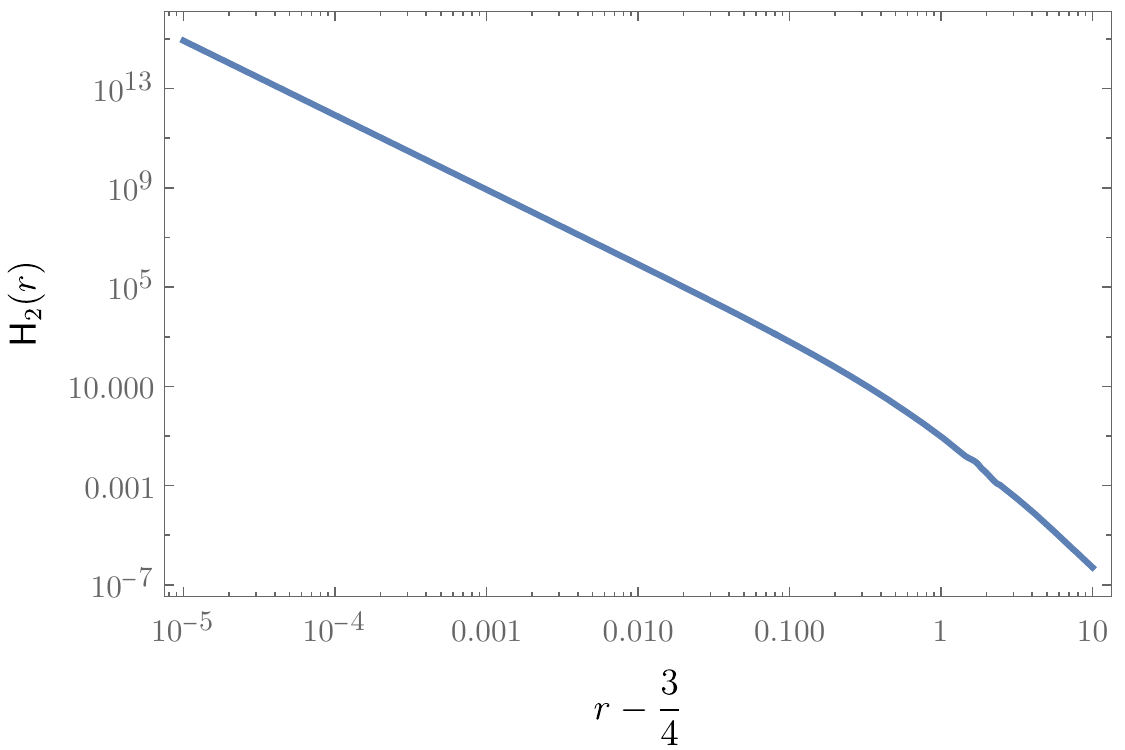}\centering
                %\end{center}
                    \caption{The behavior of the numerical solution for $\mathsf{H}_2$ near the universal horizon. The boundary condition at large $r$ is given by
                    $\mathcal{B}_0=1$ and $\mathcal{B}_2=0$.
                    The modified gravity parameter is given by $\beta=1/5$.}
                \label{fig:loglog_chi_l=2_uh.pdf}
            \end{minipage}
            \hspace{0.04\columnwidth}
            \begin{minipage}[b]{0.45\columnwidth}
                    \centering    \includegraphics[keepaspectratio=true,height=50mm]{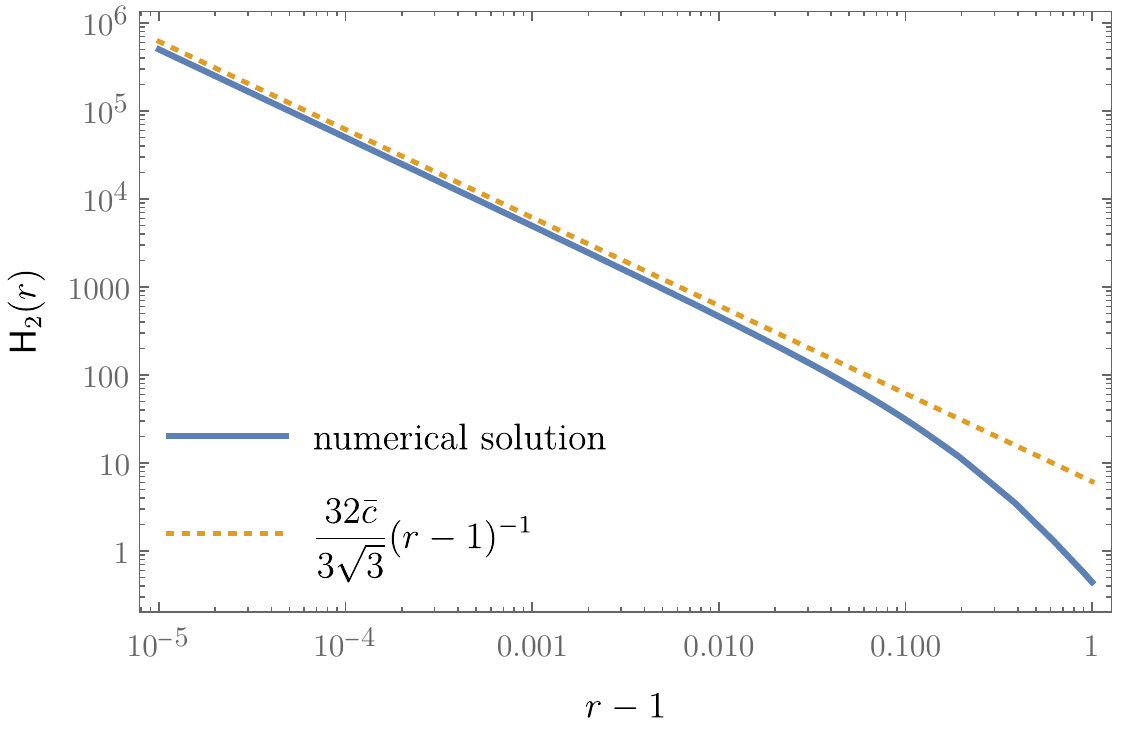}\centering
                 \caption{The behavior of the numerical solution for $\mathsf{H}_2$ near the event horizon (blue solid line).
                 The boundary condition at large $r$ is given by
                    $\mathcal{B}_0=0$ and $\mathcal{B}_2=1$.
                    The modified gravity parameter is given by $\beta=1/5$.
                    The analytic result is also shown as the orange dotted line, with $\bar c=1$.}
                \label{fig:loglog_chi_l=2_kh.pdf}
            \end{minipage}
        \end{figure}

Having determined the solution for large $r$, let us investigate the behavior of the perturbations near the inner boundaries, focusing on the case of the regular universal horizon, $b_0=b_{0,c}$.
If one sets $\mathcal{B}_2=0$ to kill the tensorial degree of freedom and integrates the equations of motion inwards, the metric perturbations are found to diverge at the universal horizon, $r=3/4$, as presented in Fig.~\ref{fig:loglog_chi_l=2_uh.pdf}.
This result shows that the terms with the coefficient $\mathcal{B}_0$ are indeed induced by the instantaneous mode, whose causal boundary is the universal horizon.
By inspecting the equations of motion in the vicinity of $r=3/4$, we find that the metric perturbations diverge as
\begin{align}
    \widetilde{\mathsf{H}}_1&\simeq \varepsilon^{1-k_\ell}(c+c'\ln\varepsilon),
    \quad  
    \mathsf{H}_2\simeq -\frac{2\sqrt{3}}{N_0}\widetilde{\mathsf{H}}_1,
    \quad 
    \mathsf{H}_0\simeq \frac{\sqrt{3}}{N_0}\varepsilon^{1-k_\ell}\left[
        c-\frac{4\sqrt{2}\beta}{N_0}\frac{(2-k_\ell)(3-k_\ell)}{\ell(\ell+1)}c'
        +c'\ln\varepsilon
        \right],\notag 
    \\ 
    \widetilde{\mathsf{a}}&\simeq -\frac{3\sqrt{3}}{8N_0(1-k_\ell)}\varepsilon^{-k_\ell}\left(
        c-\frac{c'}{1-k_\ell}+c'\ln\varepsilon
    \right),\notag 
    \\
    \widetilde{\mathsf{b}}&\simeq \frac{4}{1-k_\ell}\varepsilon^{2-k_\ell}\left[
        c-c'\left(\frac{1}{1-k_\ell}+\frac{2\sqrt{2}\beta}{N_0}\right)
        +\frac{2(1-k_\ell)(3-k_\ell)}{\ell(\ell+1)}c'\ln\varepsilon
    \right],
\end{align}
where $\varepsilon:=r-3/4$, $k_\ell:=2+\sqrt{1+\ell(\ell+1)/2}$,
and $c$ and $c'$ are constants.

If one sets $\mathcal{B}_2\neq 0$ and integrates the equations of motion inwards, one finds that the metric perturbations diverge at the usual event horizon, $r=1$, before reaching the universal horizon, as shown in Fig.~\ref{fig:loglog_chi_l=2_kh.pdf}.
Recalling that the tensorial degrees of freedom propagate at the speed of light in the present theory, it is reasonable that the inner boundary in this case is given by the usual event horizon.
In this case, one can study the behavior of the metric perturbations near the event horizon
by expanding the equations of motion around $r=1$ to obtain
\begin{align}
    \widetilde{\mathsf{H}}_1&\simeq \bar{c}(r-1)^{-1},\quad 
    \mathsf{H}_2\simeq  \frac{32\bar{c}}{3\sqrt{3}N_0}(r-1)^{-1},\quad 
    \mathsf{H}_0\simeq %\mathrm{const}
    -\frac{1184\bar{c}}{81\sqrt{3}N_0}\ln{(r-1)},\notag \\
    \widetilde{\mathsf{a}}&\simeq \frac{32\bar{c}}{3\sqrt{3}\ell(\ell+1)N_0}(r-1)^{-1},\quad 
    \widetilde{\mathsf{b}}\simeq \frac{2\bar{c}}{\ell(\ell+1)}(r-1)^{-1},
\end{align}
where $\bar c$ is a constant.
In any case the stationary perturbations diverge at the inner boundaries,
leading to the conclusion that no regular perturbations with $\ell\ge 2$ are allowed.

% %----------------------------------------------%
% %-----------------------------------------------%

%As shown in Fig.~\ref{fig:num-solution_chi_l=2.pdf} and Fig.~\ref{fig:num-solution_chi_l=2_loglog.pdf}, we can solve numerically up to the universal horizon.

\section{Conclusions}\label{conclusion}

We have considered a spatially covariant theory of gravity
having just two tensorial degrees of freedom and a non-propagating (instantaneous) scalar mode.
There is no propagating scalar mode, and therefore
the number of propagating degrees of freedom is the same as in general relativity (GR).
In a particular subset of such theories~\cite{Gao:2019twq},
the standard Newtonian behavior of gravity is reproduced
and the propagation of gravitational waves in a cosmological background obeys the same equation
as in GR~\cite{Iyonaga:2021yfv}.
Moreover, in the same subset of theories of~\cite{Gao:2019twq},
the Schwarzschild solution foliated by the maximal slices is
a solution to the field equations, as in the case of
Einstein-Aether theory~\cite{Berglund:2012bu}.
The solution forms a universal horizon,
which is the causal boundary for the scalar mode with infinite propagation speed~\cite{Iyonaga:2021yfv}.
In this paper, we have studied linear perturbations of this black hole solution.

First, we have studied the odd-parity perturbations.
The odd-parity sector contains only one of the two tensorial modes but is devoid of contributions from the instantaneous scalar mode.
We have presented the quadratic Lagrangian for the odd-parity perturbations,
which turned out to be identical to that in GR after an appropriate identification of the variables. Since the tensorial modes propagate at the speed of light, the causal boundary of the odd-parity perturbations is given by the usual even horizon rather than the universal horizon inside of it.

Next, we have considered the even-parity sector of black hole perturbations, in which the instantaneous scalar mode and one of the two tensorial modes are mixed.
We have derived quadratic actions for monopole, dipole, and higher multipole modes ($\ell\ge 2$), and studied them separately.
The tensorial mode does not contribute to the monopole and dipole perturbations.
As a result, we have found no radiative behavior in these perturbations.
We have solved the set of field equations for the monopole perturbations and found no solution that is regular both at infinity and at the inner boundary, i.e. the universal horizon, except for the trivial one corresponding to a shift of the mass parameter.
For the dipole perturbations, we have derived a single master equation which is a fourth-order differential equation with respect to the radial coordinate. Again, we have found no solution that is regular both at infinity and at the universal horizon.

The situation gets more complicated for the even-parity perturbations with $\ell\ge 2$.
We have derived the general quadratic action.
Although there is only one propagating degree of freedom
in the even-parity sector with $\ell\ge 2$,
we have not been able to reduce the system to a single master equation
because of the complexity of the equations stemming from the mixing with
the instantaneous scalar mode.
We therefore focused on stationary perturbations and investigated their properties.
We have found that the stationary perturbations with $\ell\ge 2$ at large $r$
are characterized by two integration constants,
$\mathcal{B}_2$ and $\mathcal{B}_0$, governing respectively the contributions
from the tensorial degree of freedom and the instantaneous scalar mode.
We carefully identified the locations of the appropriate inner boundaries
and showed that the perturbations diverge at the inner boundaries in any case
unless $\mathcal{B}_2=\mathcal{B}_0=0$.

To conclude, we have developed black hole perturbation theory
for the Schwarzschild solution foliated by the maximal slices
in spatially covariant gravity with just two tensorial degrees of freedom
and established its perturbative uniqueness.
As a future direction, it would be interesting to study the wave-like behavior of
the even-parity perturbations with $\ell\ge 2$ to see the impact of mixing with
the instantaneous scalar mode on gravitational waves,
which is technically more difficult and is left for further study.

%--- Acknowledgments ---%--- Acknowledgments ---%--- Acknowledgments ---%
\acknowledgments
The work of JS was supported by
the Rikkyo University Special Fund for Research.
The work of TK was supported by
JSPS KAKENHI Grant No.~JP20K03936 and
MEXT-JSPS Grant-in-Aid for Transformative Research Areas (A) ``Extreme Universe'',
No.~JP21H05182 and No.~JP21H05189.
%--- Acknowledgments ---%--- Acknowledgments ---%--- Acknowledgments ---%

%\appendix

%-----------------------------------------------------------------%
\bibliography{refs}
\bibliographystyle{JHEP}
%-----------------------------------------------------------------%
\end{document}